# Reinforcement Learning-Enabled Decision-Making Strategies for a Vehicle-Cyber-Physical-System in Connected Environment

Teng Liu
Department of Automotive Engineering and the State Key Laboratory of Mechanical Transmission
Chongqing University, Chongqing 400044, China
tengliu17@gmail.com

Xiaolin Tang
Department of Automotive Engineering and the State Key Laboratory of Mechanical Transmission
Chongqing University, Chongqing 400044, China
tangxl0923@cqu.edu.cn

Jinwei Zhang
Department of Mechanical and Mechatronics Engineering
University of Waterloo
Ontario N2L3G1, Canada
jinwei.zhang@uwaterloo.ca

Wenbo Li
Department of Automotive Engineering and the State Key Laboratory of Mechanical Transmission
Chongqing University, Chongqing 400044, China
liwenbocqu@foxmail.com

Zejian Deng
Department of Mechanical and Mechatronics Engineering
University of Waterloo
Ontario N2L3G1, Canada
z49deng@uwaterloo.ca

Yalian Yang
Department of Automotive Engineering and the State Key Laboratory of Mechanical Transmission
Chongqing University, Chongqing 400044, China
YYL@cqu.edu.cn

*Abstract*—As a typical vehicle-cyber-physical-system (V-CPS), connected automated vehicle attracted more and more attention in recent years. This paper focuses on discussing the decision-making (DM) strategy for autonomous vehicles in a connected environment. First, the highway DM problem is formulated, wherein the vehicles can exchange information via wireless networking. Then, two classical reinforcement learning (RL) algorithms, Q-learning and Dyna, are leveraged to derive the DM strategies in a predefined driving scenario. Finally, the control performance of the derived DM policies in safety and efficiency is analyzed. Furthermore, the inherent differences of the RL algorithms are embodied and discussed in DM strategies.

## I. INTRODUCTION

Cyber-physical systems (CPSs) caught more and more attention from industrial and academic communities in recent years, and they refer to the intimate combination and coordination between physical and computational resources [1]. The physical components are intertwined with software ones to construct more quickly, precise and efficient systems. A CPS could have several crucial characteristics, such as high-degree of automation, cyber capability, networking at multiple scales and reorganizing dynamics [2]. Smart grid, robotics system, autonomous automobile systems and the medical monitor can all be regarded as CPSs [3].

Connected automated vehicles (CAVs) is a representative CPS, wherein the physical components determine the motion and manage the energy consumption, and the cyber parts blend multi-sensors' information and interact with other vehicles and infrastructures. For example, the authors in [4] developed a performance evaluation model for unmanned connected vehicle cyber-physical-system (V-CPS). In order to guarantee accurate and fast communication, the particle swarm optimization (PSO) algorithm is applied to improve the positional accuracy of the wireless sensor network. Loos et al. built a distributed car control system to enhance road-safety for autonomous vehicles [5]. Cyber technologies utilized sensors data to guide the physical aspects of car movement and avoid collisions. Furthermore, to enable CAVs to operate human-like behaviors, Ref. [6] applied machine learning methods to learn the stochastic characteristics of the human driver. To verify the performance of automotive CPSs, the authors integrate the path prediction model into the algorithms to forecast the human steering preferences and improve accuracy.

Decision-making is one of the most significant modules in CAVs. It indicates generating a sequence of motion actions for physical components of CPS to satisfy particular functions requirements, such as lane changing, merging into the highway and traversing an unprotected intersection. For example, the authors in [7] reviewed popular approaches for decision-making in CAVs, such as probabilistic algorithms, partially observable Markov decision process (POMDP), reinforcement learning (RL) and inverse RL. Galceran et al. proposed a multipolicy decision making for the automated vehicle via changepoint-based behavior prediction. Simulated and real-world experiments were constructed to evaluate the predicted outcomes and decision-making strategy. Moreover, Ref. [9] focused on the decision-making behavior at uncertain interaction and employed POMDP to model the driving intention of surrounding vehicles. An interactive, probabilistic motion model was used to forecast the future actions of other vehicles and help the ego vehicle to make suitable and right decisions.

Recently, learning-based methods are regarded as potential solutions for intelligent and advanced decision-making policies for CAVs [10-12]. In this paper, an RL-enabled freeway overtaking decision-making strategy is presented for V-CPS in connected environments. First, the problem formulation is stated, wherein the driving scenario for highway overtake is described. In this situation, each vehicle could exchange position and velocity information with others. Then, two popular RL algorithms, Q-learning and Sarsa are introduced and the theoretical differences between them are addressed. Finally, the effects on the safety and efficiency of these two methods are compared and analyzed. The preferences and future development of overtaking policies on the highway are also specified and outlook.

The construction of the rest of the paper is organized as follows: the construction of the discussed problem is given in Section II. The learning-based methods are described in Section III, wherein the elements of RL algorithms are shown in detail. Simulation results are evaluated and analyzed in Section IV, and Section V concludes the paper.

## II. PROBLEM STATEMENT

In the content of this section, the discussed freeway overtake problem is built. Safety and efficiency concerns are explained here. Also, the evolution of vehicle speed and position of ego and surrounding vehicles are described. Finally, the optimization objective of this founded problem is formulated

### A. Driving Scenario

In CAVs, the decision-making usually refers to decide the motion of lateral and longitudinal directions. For the control action layers, the acceleration and steering angle are the key features. The typical driving scenarios for decision-making in autonomous vehicles are highway overtaking and intersection turning [13, 14]. The objective vehicle is often named as ego vehicle and the nearby vehicles are called as surrounding vehicles. The ego vehicle should make efficient decisions while considering the driving intention and behaviors of surrounding vehicles.

This work focuses on developing decision-making policies for two-lane highway overtaking problem. As depicted in Fig. 1, the brown vehicle is ego vehicle and the white vehicles are surrounding vehicles. The lanes in this situation are labeled as $L$=1 or 2, and all the vehicles are assumed to run in the same direction. The number of surrounding vehicles is 2n, which means there are n vehicles in each lane. The ego vehicle can make timely and appropriate decisions to realize lane change and surpass the surrounding vehicles.

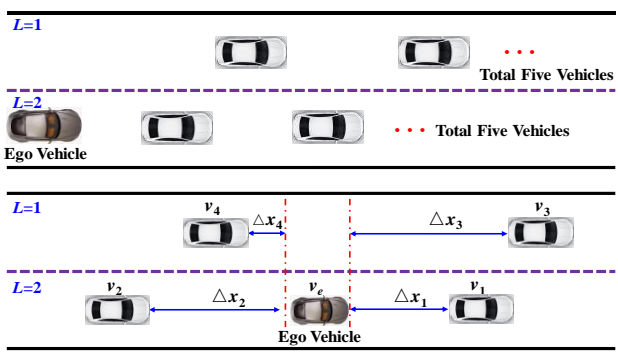

Fig. 1. Two lanes freeway overtaking scenario and the related CAVs.

### B. Driving States and Control Actions

In this decision-making problem, the ego vehicle aims to run from the start position to the final position as soon as possible without colliding other vehicles. The travel time is interpreted as efficiency and the collision cases are the safety factors. The position and velocity of the surrounding vehicle are randomly defined, and this information can transmit to the ego vehicle by the wireless network.

Based on this communication information, the ego vehicle only treats the former and latter vehicle in each lane as surrounding vehicles. Therefore, at each time step, the number of surrounding vehicles would not exceed 4. The relative velocity and distance are regarded as the state variables in this freeway overtaking problem

$$\begin{cases} \Delta x_i = |x_i - x_e| \\ \Delta v_i = |v_i - v_e| \end{cases}, i=1, 2, 3, 4 \qquad (1)$$

where $\Delta x$ and $\Delta v$ are the relative distance and velocity. $v_e$ and $x_e$ are the speed and position of the ego vehicle, same as the surrounding vehicles. They are computed as follow

$$\begin{cases} x_i = [v_i^2 - (v_i')^2]/2a_i \\ a_i = [v_i - v_i']/\Delta t \end{cases}, i=1, 2, 3, 4 \qquad (2)$$

$$\begin{cases} v_e = v_e' + a_e \cdot \Delta t \\ x_e = v_e' \cdot \Delta t + \frac{1}{2} a_e \cdot \Delta t^2 \end{cases} \qquad (3)$$

where $a_i$ and $a_e$ are the accelerations for surrounding and ego vehicles, respectively. $v_e'$ and $v_i'$ are their speed at the previous time step. $\Delta t$ is the sampling time step. It should be noticed that the vehicle velocity of the surrounding vehicles is randomly predefined, and thus their position and acceleration could be calculated in (2).

For the ego vehicle, the control actions need to be decided at each time step are the acceleration and lane number $L$. They belong to $a_e \in \{-1, 0, 1\}$ and $L \in \{1, 2\}$. The feasible control actions are listed in Table I. Finally, the objective of this problem has two concerns, safety, and efficiency. The former means the collision avoidance is the leading position, and then the latter requires the ego vehicle to pass the driving scenario as soon as possible. The mathematic expression of the control goal is given in the next section, which is represented as the reward function in the RL framework.

TABLE I
FEASIABLE CONTROL ACTIONS FOR EGO VEHICLES

| Label | Description |
|---|---|
| $a_1$ | Lane 1, Acceleration |
| $a_2$ | Lane 1, Deceleration |
| $a_3$ | Lane 1, Maintain speed |
| $a_4$ | Lane 2, Acceleration |
| $a_5$ | Lane 2, Deceleration |
| $a_6$ | Lane 2, Maintain speed |

## III. Q-LEARNING AND SARSA ALGORITHMS

Two representative RL algorithms, Q-learning and Sarsa algorithms are introduced in this section. They are employed to learn the decision-making strategies for the constructed scenario in Section II. The elements in the RL framework, e.g., state variables, control actions, reward and transition model are explained detailly. The theoretical differences between these two algorithms are also described.

### A. Markov Decision Process

RL describes an intelligent agent learns to achieve a goal by interacting with the environment [15]. The learned information from the environment is known as experiences. More experiences denote more accurate derived control actions. Based on the knowledge of the environment is known or not, RL algorithms can be divided into two categories, model-free and model-based [16]. Model-based algorithms indicate the model of the environment should be built first, and then it can be used to find the optimal strategy. Model-free algorithm means the mentioned model is not necessary and the agent need to spend more time to collect experiences.

Markov decision processes (MDPs) are usually employed to mimic the RL framework, in which the chosen actions not only affect immediate reward but also the future ones. In RL, MDP is expressed as a quintuple $<S, A, P, R, \beta>$, wherein $s \in S$ and $a \in A$ are the state variable and control action, respectively. $p \in P$ and $r \in R$ are the transition model for state $s$ and reward model for state-action pair $(s, a)$, and $\beta$ is a discount factor to balance the importance of immediate and future rewards.

As defined above, the state variables are the relative speed and distance. Hence, it can be expressed as $S=\{X_1, …, X_4, V_1, …, V_4\}$. $X_i$ and $V_i$ are the indexes of the relative distance and speed and they are computed as

$$D_i = round(\Delta x_i * N_D / x_{max}), i = 1,...4 \quad (4)$$
$$V_i = round(\Delta v_i * N_V / v_{i,max}), i = 1,...4 \quad (5)$$

where *round* means rounding function, and $N_D$ and $N_V$ are the total numbers of the corresponding index. $x_{max}$ is the maximum detection distance of the ego vehicle, which implies that the relative distance is $x_{max}$ when it is greater than this value. $v_{i,max}$ is the maximum speed of each surrounding vehicles.

Furthermore, the control action set $A$ is depicted in Table I. The transition model $P$ of the state variables are given by (1)-(3). The reward model $R$ is shown as

$$\begin{cases} r = -100, & collision = 1 \\ r = -10 \cdot (v_e - v_{e,max})^2, & collision = 0 \end{cases} \quad (6)$$

where *collision* =1 signifies that the ego vehicle crashed other vehicles or edge of the road, and $v_{e,max}$ is the maximum velocity of ego vehicle.

### B. Q-learning and Sarsa Algorithms

In different RL algorithms, the identical motivation is finding a sequence of control actions (named as control policy $\pi$) to maximize the expected cumulative rewards, which is known as the value function. Two kinds of value functions are generally leveraged in RL

$$V^\pi(s_t) = E\{\sum_t^\infty \beta^t r_t | s_t\} \quad (7)$$

$$Q^\pi(s_t, a_t) = E\{\sum_t^\infty \beta^t r_t | s_t, a_t\} \quad (8)$$

where they are named as state value function and action-value function. $s_t$ and $a_t$ are the initial state and action. The difference between these two functions is that the current action is known or not. Thus, the action value function is often used to derive the optimal action at each step. To realize this, (8) could be rewritten as the Bellman form

$$Q^\pi(s_t, a_t) = r(s_t, a_t) + \beta \sum_{s_{t+1} \in S} p_{s_t, s_{t+1}} Q^\pi(s_{t+1}, a_{t+1}) \quad (9)$$

where $p_{st, st+1}$ denotes the transition probability from state $s_t$ to $s_{t+1}$, and $a_{t+1}$ is the action related to state $s_{t+1}$. Then, the optimal control action is obtained via searching the maximum value function

$$\pi^*(s_t) = \arg\max_{a_t} [r(s_t, a_t) + \beta \sum_{s_{t+1} \in S} p_{s_t, s_{t+1}} Q^\pi(s_{t+1}, a_{t+1})] \quad (10)$$

It is obvious that the optimal control policy is completely determined by the action value matrix $Q(s, a)$. Moreover, the $Q(s, a)$ is decided by the control action at each time step. The $\varepsilon$-greedy policy is usually used to select control actions. It represents that the agent explores a random action with probability $\varepsilon$ to increase the experiences, and exploit the best action in the $Q(s, a)$ matrix until now with probability 1-$\varepsilon$.

In RL algorithms, if the current and next actions are all chosen by $\varepsilon$-greedy policy, this algorithm is called on-policy. Oppositely, if the next action is not selected by $\varepsilon$-greedy policy, it is named off-policy. The typical off-policy and on-policy algorithms are Q-learning and Sarsa. Their updating equation are depicted as follows [17, 18]

$$Q(s,a) \leftarrow Q(s,a) + \alpha[r + \beta \max_{a'} Q(s',a') - Q(s,a)] \quad (11)$$

$$Q(s,a) \leftarrow Q(s,a) + \alpha[r + \beta Q(s',a') - Q(s,a)] \quad (12)$$

where $\alpha \in [0, 1]$ is a learning rate to trade-off the old and new learned knowledge. $s'$ and $a'$ are the next state and action. The next action in (11) is selected via maximum probability and it is chosen by $\varepsilon$-greedy policy in (12). $\varepsilon$-greedy policy indicates spending more time to collect experiences and found the environment model. Maximum probability means selecting the next action depends on the current best knowledge. Hence, the Sarsa may consume more time, however, its performance may better than Q-learning.

These two RL algorithms are realized in Matlab through MDP toolbox [19]. The arguments in these algorithms are determined after a series of trials, wherein the learning rate $\alpha$ and discount factor $\beta$ are 0.9 and 0.2, respectively. The probability $\varepsilon$ is equal to $0.1*0.99^t$ and decreases with the time steps. The episode number is 200 and each episode contains 1000 steps.

## IV. RESULTS AND DISCUSSION

This section discusses the control performance of two RL algorithms for highway overtaking problem. The collision conditions, travel distance, and control actions are compared.

Based on the results, suitable algorithms are recommended in different driving scenarios.

### A. Comparison of Two Algorithms

In this subsection, Q-learning and Sarsa are compared in the same driving scenario. The number of surrounding vehicles on each lane is n=5. The parameters in these two situations are the same, such as the learning rate, maximum speed, and discrete index.

The collision cases in these two algorithms are shown in Fig. 2. It is obvious that Q-learning requires more time to learn how to avoid the collision. In the beginning, the Q-learning cannot reach the final point (the distance is 1000 m) because it may maintain the speed as zero. For Sarsa, it can learn very quickly and reach the final point. However, sometimes the collision is equal to 1 again due to the $\varepsilon$-greedy policy. Thus, if the driving scenario is simple enough, the Sarsa can solve it. Oppositely, Q-learning is able to learn the more complex overtaking problem.

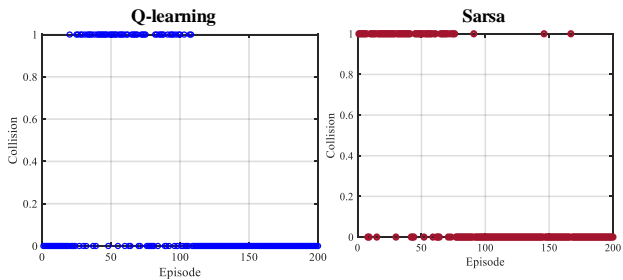
Fig. 2. Collision conditions in two RL algorithms for the same parameters.

For the different episode, the travel distance and consumed time in these two cases are depicted in Fig. 3. It can be discerned that Sarsa requires fewer episodes to reach the final point and it would cost more time. This is caused by the $\varepsilon$-greedy policy that it will choose the next action randomly. Hence, Sarsa is suitable to be applied in an easy situation, wherein the consumed time would not be too long.

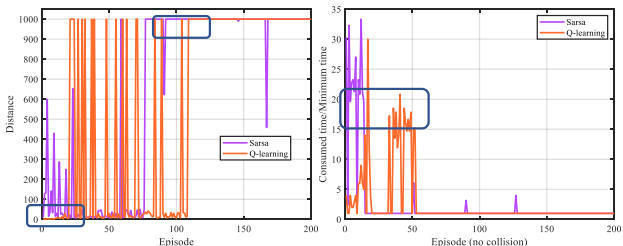
Fig. 3. Travel distance and consumed time in Sarsa and Q-learning.

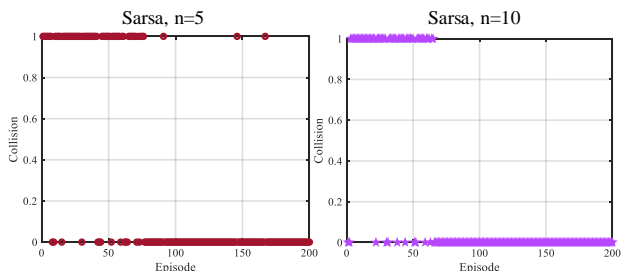
Fig. 4. Collision conditions in Sarsa for different surrounding vehicles.

### B. Parameters influence in Sarsa

After analyzing the control performance of different RL algorithms, the effects of parameters for one algorithm are discussed in this subsection. First, the number of surrounding vehicles are changed are n=5 and n=10. The collision conditions are described in Fig. 4. As more vehicles exist the scenario, the ego vehicle could learn more quickly, but the consumed time will be longer. The relevant control actions are displayed in Fig. 5, wherein they are more complicated in the second driving scenario.

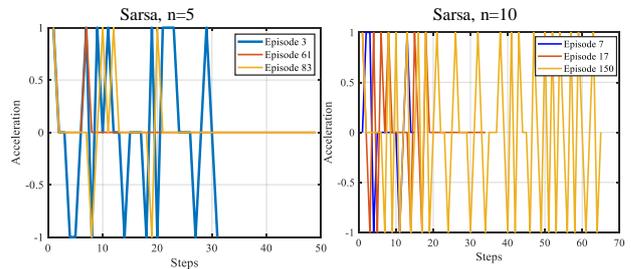
Fig. 5. Acceleration in Sarsa for different surrounding vehicles.

As discussed above, the value of $\varepsilon$ is very important for RL algorithms. To evaluate its influence on the convergence rate, it is assigned to different values. The first one is $\varepsilon=0.1*0.9^t$ and the second one is $\varepsilon=0.1$. The collision conditions in these two cases are given in Fig. 6. It is very apparent that for the fixed value of $\varepsilon$, the convergence rate is very slow and it cannot learn to avoid the collision in 200 episodes. Inversely, if the $\varepsilon$ decreases with the time steps, it will improve the learning speed. Hence, how to decide the value of selection probability $\varepsilon$ is significant for RL-based decision-making strategies for automated vehicles.

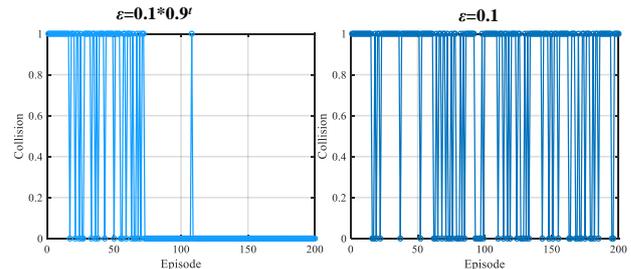
Fig. 6. Collision conditions for different $\varepsilon$ values in Sarsa algorithm.

### V. CONCLUSION

The RL-enabled decision-making strategy is formulated in this paper to manage the autonomous vehicle complete free-way overtaking task. The driving environment is first introduced and the definition of ego vehicle and surrounding vehicles are given. Then, the Q-learning and Sarsa algorithms are described, wherein the theoretical differences between these two algorithms are explained. The merits of demerits of these two algorithms on highway overtaking problem are illuminated. Future work focuses on combining deep learning and improved RL algorithm to derive online decision-making strategies for autonomous vehicles.